%% file: main.tex
\documentclass[runningheads]{llncs}
\usepackage[T1]{fontenc}
\usepackage{graphicx}
\usepackage{booktabs}
\usepackage[misc]{ifsym}
\newcommand{\corr}{(\Letter)}

\usepackage{latexsym}
\usepackage{amsfonts}
\usepackage{amsmath}
\usepackage{subfigure}
\usepackage{mathtools}
\usepackage{relsize}
\usepackage{dsfont}
\usepackage{enumitem}

\usepackage{booktabs}
\usepackage{pifont}
\newcommand{\cmark}{\text{\ding{51}}}
\newcommand{\xmark}{\text{\ding{55}}}

\DeclareMathOperator*{\argmax}{arg\,max}
\DeclareMathOperator*{\argmin}{arg\,min}

\usepackage{algorithm2e}
\SetKw{Input}{Input}
\SetKw{Output}{Output}
\RestyleAlgo{ruled}
\usepackage{multirow}
\usepackage{booktabs}
\usepackage{adjustbox}
\usepackage{wrapfig}
\newcommand{\rulesep}{\unskip\ \vrule\ }

\begin{document}

\title{PromptDSI: Prompt-based Rehearsal-free Continual Learning for Document Retrieval}
\toctitle{PromptDSI: Prompt-based Rehearsal-free Continual Learning for Document Retrieval}

\titlerunning{PromptDSI}


\author{
    Tuan-Luc Huynh\inst{1} \corr \and 
    Thuy-Trang Vu\inst{1} \and 
    Weiqing Wang\inst{1} \and 
    Yinwei Wei\inst{2} \and
    Trung Le\inst{1} \and 
    Dragan Gasevic\inst{2} \and 
    Yuan-Fang Li\inst{1} \and 
    Thanh-Toan Do\inst{1}
}

\tocauthor{
    Tuan-Luc Huynh, 
    Thuy-Trang Vu, 
    Weiqing Wang, 
    Yinwei Wei,
    Trung Le, 
    Dragan Gasevic, 
    Yuan-Fang Li, 
    Thanh-Toan Do
}

\institute{ 
Department of Data Science \& AI, Monash University, Australia \email{tuan.huynh1@monash.edu} \and 
Department of Human-Centred Computing, Monash University, Australia
}

\authorrunning{TL Huynh et al.}


\maketitle              

\input{0_abstract}
\input{1_introduction}
\input{2_related_work}
\input{3_preliminary}
\input{4_promptDSI}
\input{5_experiments}
\input{6_conclusion}

\begin{credits}
\subsubsection{\ackname} 
This material is based on research sponsored by Defense Advanced Research Projects Agency (DARPA) under agreement number HR0011-22-2-0047. The U.S. Government is authorised to reproduce and distribute reprints for Governmental purposes notwithstanding any copyright notation thereon. The views and conclusions contained herein are those of the authors and should not be interpreted as necessarily representing the official policies or endorsements, either expressed or implied, of DARPA or the U.S. Government.
This work is supported by the Australian Research Council Discovery Early Career Researcher Award DE250100032 
and Monash eResearch capabilities, including M3.

\end{credits}
\bibliographystyle{splncs04}

\end{document}

%% file: 0_abstract.tex
\begin{abstract}
Differentiable Search Index (DSI) utilizes pre-trained language models to perform indexing and document retrieval via end-to-end learning without relying on external indexes. However, DSI requires full re-training to index new documents, causing significant computational inefficiencies. Continual learning (CL) offers a solution by enabling the model to incrementally update without full re-training. Existing CL solutions in document retrieval rely on memory buffers or generative models \textit{for rehearsal}, which is infeasible when accessing previous training data is restricted due to privacy concerns.
To this end, we introduce PromptDSI, a prompt-based, \textit{rehearsal-free} continual learning approach for document retrieval. PromptDSI follows the Prompt-based Continual Learning (PCL) framework, using learnable prompts to efficiently index new documents without accessing previous documents or queries. To improve retrieval latency, we remove the initial forward pass of PCL, which otherwise greatly increases training and inference time, with a negligible trade-off in performance. 
Additionally, we introduce a novel topic-aware prompt pool that employs neural topic embeddings as fixed keys, eliminating the instability of prompt key optimization while maintaining competitive performance with existing PCL prompt pools. In a challenging rehearsal-free continual learning setup, we demonstrate that PromptDSI variants outperform rehearsal-based baselines, match the strong cache-based baseline in mitigating forgetting, and significantly improving retrieval performance on new corpora\footnote{Code is available at: \url{https://github.com/LouisDo2108/PromptDSI}.}.
\end{abstract}

%% file: 1_introduction.tex
\section{Introduction}
Differentiable Search Index (DSI)~\cite{tay_dsi_neurips22} leverages a Transformer model~\cite{vaswani_attention_neurips17} to encode corpus information directly into the model parameters through end-to-end optimization. This end-to-end retrieval paradigm eliminates the need to construct traditional inverted indices or vector databases used in sparse~\cite{robertson_bm25_2009} or dense retrieval~\cite{karpukhin_dpr_emnlp20}.
However, in real-world scenarios where new documents are continually added, re-training DSI from scratch for each update is computationally prohibitive, necessitating continual learning (CL) methods~\cite{mccloskey_cl_1989,french_lifelong_1999}. 
This challenging setting is known as dynamic corpora~\cite{mehta_dsipp_emnlp23}, or lifelong information retrieval~\cite{gerald_ircl_ecir22}.
DSI++~\cite{mehta_dsipp_emnlp23} employs generative replay, while IncDSI~\cite{kishore_incdsi_icml23} uses constrained optimization in an instance-wise continual learning setup. However, DSI++ requires an additional query generation model, leading to substantial computational overhead during training and introducing the challenge of maintaining this model for continual indexing. Meanwhile, IncDSI relies on caching queries for all previously indexed documents, resulting in high memory demands and potential data privacy concerns~\cite{custers_privacy_2019}.

Rehearsal-free prompt-based continual learning (PCL) methods~\cite{wang_l2p_cvpr22,wang_dualprompt_eccv22,wang_sprompt_neurips22,smith_codaprompt_cvpr23,wang_hideprompt_neurips23} offer a promising approach to alleviate the need to access previous documents or queries. These methods have shown competitive performance compared to rehearsal-based techniques in class-continual learning settings within the vision domain~\cite{wang_hideprompt_neurips23}. 
However, adapting PCL methods to document retrieval remains unexplored due to the extreme classification problem caused by the instance-wise nature, in which each document is a unique class and there are usually at least 100k documents.
Moreover, existing PCL methods require two forward passes through the Transformer model, which is not suitable for retrieval systems with low latency requirements.

In this work, we propose PromptDSI, a prompt-based \emph{rehearsal-free} continual learning DSI for document retrieval. PromptDSI uses learnable prompts to index new documents while keeping the the DSI backbone frozen, while not accessing previous documents or queries. 
To overcome the inefficiencies of PCL methods and tailor them for retrieval, we introduce several model-agnostic modifications.
We eliminate the inefficient initial forward pass in PCL methods by using intermediate layer representations instead of the pre-trained language model's final layer \texttt{[CLS]} token for query-key matching. This reduces computational overhead with a minimal performance trade-off.
Furthermore, due to the lack of distinct semantics across new corpora, PromptDSI often collapses to using a limited set of prompts, leading to underutilized parameters. Inspired by neural topic modeling~\cite{gupta_topic_icml20,gerald_ircl_ecir22}, we propose using neural topic embeddings mined from the initial corpus as fixed keys in the prompt pool of PCL methods. This strategy eliminates the training instability of prompt keys and improves the interpretability of prompt selection.
Finally, while existing PCL methods follow multi-layer prompting~\cite{wang_dualprompt_eccv22,smith_codaprompt_cvpr23,wang_hideprompt_neurips23}, it is unclear if this is optimal for document retrieval. To explore the stability-plasticity trade-off, we conduct a comprehensive layer-wise prompting study to identify the most effective prompting layers. Overall, our work makes the following contributions:
\begin{itemize}
    \item We introduce PromptDSI, the first PCL method 
    for classification-based, end-to-end document retrieval.
    \item We propose two novel approaches to adapt existing PCL methods to document retrieval: (i) an efficient single-pass PCL approach for low-latency retrieval and (ii) using neural topic embeddings as fixed prompt keys to stabilize query-key matching in PCL's prompt pool, addressing prompt underutilization and enhancing interpretability.
    \item We conduct a thorough analysis which verifies that single-layer prompting is sufficient for optimal performance when adapting existing PCL methods to continual learning for document retrieval.
    \item Experimental results on the NQ320k and MSMARCO 300k datasets under the challenging rehearsal-free setup show that PromptDSI performs on par with IncDSI, a strong baseline that requires caching previous training data.
\end{itemize}

%% file: 2_related_work.tex
\section{Related Work}
\subsection{End-to-end Retrieval}
End-to-end retrieval is an emerging paradigm that aims to replace the conventional ``retrieve-then-rank'' pipeline~\cite{robertson_bm25_2009,karpukhin_dpr_emnlp20} with a single Transformer model~\cite{vaswani_attention_neurips17}, pioneered by Differentiable Search Index (DSI)~\cite{tay_dsi_neurips22}. By integrating corpus information into the model parameters, DSI eliminates specialized search procedures, and enables end-to-end optimization.
DSI can be divided into two subcategories: classification-based~\cite{kishore_incdsi_icml23} and generative retrieval~\cite{mehta_dsipp_emnlp23,chen_clever_cikm23}, with the former being more resilient to catastrophic forgetting during continual indexing~\cite{mehta_dsipp_emnlp23}. Therefore, we focus on the classification-based DSI approach in this work.
Improving document identifier representation remains the main research focus of the community~\cite{zeng_ripor_www24,zhang_tsgen_sigir24}, while the challenging continual learning (CL) setting is receiving increasing attention~\cite{mehta_dsipp_emnlp23,kishore_incdsi_icml23,chen_clever_cikm23}. However, existing solutions either rely on caching previous queries or a generative model for rehearsal. PromptDSI builds upon IncDSI~\cite{kishore_incdsi_icml23}, which handles CL by caching all previous queries, and pushes DSI towards rehearsal-free CL.

\subsection{Continual Learning (CL)}
Addressing catastrophic forgetting in continual learning (CL) has been an active research area~\cite{wang_clsurvey_23}. Two popular approaches are regularization-based methods, which constrain model updates via regularization terms or knowledge distillation~\cite{kirkpatrick_ewc_pnas17,li_lwf_tpami17}, and replay-based methods, which use memory buffers or generative models for rehearsal~\cite{chaudhry_agem_iclr19,buzzega_der_neurips20}. These approaches dominate lifelong information retrieval studies~\cite{mehta_dsipp_emnlp23,chen_clever_cikm23}. Recently, Prompt-based Continual Learning (PCL) has emerged as a solution for scenarios where historical data access is restricted due to privacy regulations~\cite{custers_privacy_2019}. Various studies~\cite{wang_l2p_cvpr22,wang_dualprompt_eccv22,smith_codaprompt_cvpr23,wang_sprompt_neurips22} have explored the use of prompts to guide frozen pre-trained models in learning new tasks without memory buffers. Concurrent to our work,~\cite{kim_osprompt_eccv24} introduces one-stage prompt-based continual learning; however, their method is applied only to CODA-Prompt~\cite{smith_codaprompt_cvpr23} and evaluated exclusively in the vision domain. In contrast, PromptDSI adapts vision-domain PCL methods to a more challenging instance-wise continual learning setting, where each document is a unique class, with at least 100k documents. We propose a model-agnostic modification: efficient single-pass PCL, and validate its performance across three popular PCL methods. Additionally, we introduce fixed neural topic embeddings as prompt keys to mitigate prompt pool underutilization observed in certain PCL methods and to enhance explainability.

%% file: 3_preliminary.tex
\section{Background}
\subsection{Differentiable Search Index (DSI)} 

\paragraph{Indexing Stage.} 
A Transformer model~\cite{vaswani_attention_neurips17} $f_\Theta$ parameterized by $\Theta$ is trained to learn a mapping function from a document $d \in \mathcal{D}$ to its document identifier (docid) $id \in \mathcal{I}$ during the indexing stage: $f_\Theta \colon \mathcal{D} \to \mathcal{I} $. \textbf{Document representation} and \textbf{docid representation} are two crucial aspects of DSI that need to be predefined.

\paragraph{Document Representation.} 
The original DSI is trained to map a document's texts to its corresponding docid. However, the model may need to process queries that are distributionally different from training documents at retrieval time, leading to a mismatch between indexing and retrieval. To bridge this gap, the current standard approach is to generate pseudo-queries that serve as document representations during indexing~\cite{zhuang_dsiqg_22,pradeep_genirscaling_emnlp23}. Following this, we leverage \texttt{docT5query}~\cite{nogueira_doct5query_2019} to generate pseudo-queries that supplement annotated queries, which are collectively used as inputs during the indexing stage: $f_\Theta \colon \mathcal{Q} \to \mathcal{I}$.

\paragraph{Docid Representation.} 
We adopt atomic docids, where each document is assigned a unique integer identifier. 
Denote the classification-based DSI model as $f_\Theta$, parameterized by $\Theta=\{\theta_{e}, \theta_{l}\}$, where $\theta_{e}$ represents the encoder weights, and $\theta_l \in \mathbb{R}^{\text{dim} \times |\mathcal{D}|}$ denotes the linear classifier weights, with $|\mathcal{D}|$ being the total number of documents. Each docid corresponds to a unique column vector $V \in \mathbb{R}^{\text{dim}}$ in $\theta_l$, where the docid (i.e., the unique integer) specifies the position of the column vector in \(\theta_l\). 
Unlike dense retrieval, which uses dual encoders and an external index with contrastive learning~\cite{karpukhin_dpr_emnlp20,xiong_ance_arxiv20}, classification-based DSI uses a single encoder and stores document embeddings as classifier weights, training with a seq2seq objective~\cite{sutskever2014sequence} to map queries to docids.

\paragraph{Retrieval Stage.}
Given a query $q \in \mathcal{Q}$, DSI returns a ranked list of documents by sorting the inner products of the query embedding $h_q=f_{\theta_e}(q)[0] \in \mathbb{R}^{\text{dim}}$ (i.e., the \texttt{[CLS]} token) and the linear classifier weight in descending order:
\begin{equation*}
    \label{eq:classfication-dsi}
    f_\Theta(q)=\left[\argmax\limits_{id\in \mathcal{I}}^{(1)}(\theta_{l}^{\top}\cdot h_q), \argmax\limits_{id\in \mathcal{I}}^{(2)}(\theta_{l}^{\top}\cdot h_q),\ldots \right],
\end{equation*}
where $\argmax\limits_{id\in \mathcal{I}}^{(i)}(\theta_{l}^{\top}\cdot h_q)$ denotes the $i\textsuperscript{th}$ ranked docid.

\subsection{Continual Learning in DSI}
\label{subsec:cl-dsi}
The continual learning setup in DSI assumes that there are $T{+}1$ corpora: $\{D_0, {\ldots}, D_T\}$ with corresponding query sets $\{Q_0, {\ldots}, Q_T\}$ and corresponding docid sets $\{I_0, {\ldots}, I_T\}$, where $D_t{=}\{d^t_1, {\ldots}, d^t_{|D_t|}\}$. $D_0$ is often a large-scale corpus with annotated query-document pairs. $D_{>0}{=}\{D_1, {\ldots}, D_T\}$ are new corpora with completely new documents and no annotated query-document pairs arriving sequentially. Denote the parameters of the DSI model at timestep $t$ (i.e., after indexing $D_{\leq t}{=}\{D_0, {\ldots}, D_t\}$) as $\Theta_t$. At every timestep $t{>}0$, the previous DSI model $f_{\Theta_{t-1}}$ trained upto $D_{\leq t-1}$ has to train on corpus $D_t$ and updates its parameters to $\Theta_t$. The evaluation at timestep $t$ will use $f_{\Theta_t}$ on $\{Q_0, {\ldots}, Q_t\}$ to predict $\{I_0, {\ldots}, I_t\}$. To index new documents, at timestep $t$, DSI's linear classifier $\theta_{l}$ is expanded to $\theta_{l}=\{\theta_{l}; \theta_{l, t}\}$ where $\theta_{l, t} \in \mathbb{R}^{\text{dim}\times |D_t|}$ denotes the expanded portion and $\{\cdot;\cdot\}$ is concatenation.

\subsection{Prompt Tuning} 
Prompt tuning~\cite{lester_ptuning_21,li_prefixtuning_21} introduces a small set of learnable soft prompts to instruct frozen pre-trained language models on downstream tasks. In this work, we adopt the prefix tuning~\cite{li_prefixtuning_21} variant.
Given an input sequence $x \in \mathbb{R}^{\text{dim}\times n}$ with length $n$ and embedding dimension of $\text{dim}$, prompt token $p \in \mathbb{R}^{\text{dim}\times m}$ with length $m$ is prepended to the input of self-attention~\cite{vaswani_attention_neurips17} as follows: 
\[\text{Attention}(xW_q, [p_k, x]W_k, [p_v, x]W_v),\] where \(W_q, W_k, W_v\) are projection matrices and $p_k, p_v \in \mathbb{R}^{\text{dim}\times\frac{m}{2}}$ are equally split from $p$.

%% file: 4_promptDSI.tex
\section{PromptDSI: Single-pass Rehearsal-free Prompt-based Continual Learning for Document Retrieval}

PCL methods~\cite{wang_l2p_cvpr22,wang_sprompt_neurips22,wang_hideprompt_neurips23,smith_codaprompt_cvpr23} can be seamlessly integrated into classification-based DSI. However, such an approach is not efficient for a document retrieval system. Moreover, simple adaptation also faces the issue of prompt underutilization. In this section, we will further elaborate on these issues and propose corresponding solutions.

\input{overview-promptdsi}

\subsection{Prompt-based Continual Learning (PCL)}
\label{subsec:pcl}
PCL methods introduce a prompt pool $\mathbf{P}=\{p_1,\ldots, p_M\}$ with $M$ prompts 
and a set of corresponding prompt keys $\mathbf{K}=\{k_1,\ldots, k_M\}$, where each prompt $p_i \in \mathbb{R}^{\text{dim}\times m}$ with length $m$ is paired with a key $k_i \in \mathbb{R}^{\text{dim}}$ in a key-value manner. 
\(\mathbf{P}\) acts as an external memory for the pre-trained language model (PLM), enabling storage of new information without disrupting its inherent knowledge or explicitly retaining previous training data.

As depicted in Figure~\ref{fig:promptdsi-architecture} (left), given a query $q$, the \textbf{first pass} extracts the \texttt{[CLS]} token $h_q$ of the input query $q$ from a frozen PLM $\theta_e$. A set of top-$N$ prompt keys from the prompt pool are optimized with cosine distance to align them with the \texttt{[CLS]} token $h_q$ using the following query-key matching mechanism:
\begin{equation}
\begin{gathered}
    \label{eq:query_key_loss}
    \mathcal{L}_{\text{match}}=\sum_{i\in S_q} \gamma(h_q, k_i), 
    \quad \text{s.t.}\quad S_q=\argmin_{\{s_i\}_{i=1}^{N}\subseteq[1, M]} \sum_{i=1}^{N} \gamma(h_q, k_{s_i}),
\end{gathered}
\end{equation}
where $S_q$ denotes a set of top-$N$ prompt key ids and $\gamma(\cdot,\cdot) = 1 - \cos(\cdot,\cdot)$ is the cosine distance. In the \textbf{second pass}, the same query \(q\) is reprocessed by \(\theta_e\); however, this time the previously selected top-$N$ corresponding prompts \(p \subset \mathbf{P}\) are prefix-tuned to the PLM, resulting in an enhanced \texttt{[CLS]} token: $\hat{h}_q = f_{p,\theta_e}(q)[0] \in \mathbb{R}^{\text{dim}}$,
which has been instructed by the knowledge from prompts \(p\).

\subsection{Single-pass PCL}
\label{subsec:osprompt-dsi}
As previously discussed, existing PCL methods involve two forward passes during training and inference. 
Given that a forward pass in Transformer models is already computationally expensive, this two-pass design is unsuitable for retrieval systems, as it exacerbates latency during both training and inference. We refer to this naive adaptation of two-pass PCL methods to DSI as Naive-PromptDSI.

To mitigate the inefficiency of Naive-PromptDSI, we introduce PromptDSI, a streamlined variant that eliminates the first pass typically required by PCL methods. As illustrated on the right side of Figure~\ref{fig:promptdsi-architecture}, instead of using the \texttt{[CLS]} token from the first pass for prompt selection, PromptDSI leverages the average token embeddings \texttt{[AVG]} from the intermediate layer immediately preceding the prompting layer: $\texttt{[AVG]} = \frac{1}{|q|} \sum_{i=1}^{|q|} f_{\theta_{e}^{l-1}}(q)_i \in \mathbb{R}^{\text{dim}}$,
where $|q|$ denotes the sequence length of query $q$, and $f_{\theta_{e}^{l-1}}(q)_i$ represents the output of the $(l-1)\textsuperscript{th}$ layer of the encoder $\theta_e$ at position $i$.
If $l=1$, we use the average of the embeddings from the pre-trained language model's embedding layer.
This adjustment approximates the semantic richness typically provided by the \texttt{[CLS]} token. 
In section~\ref{subsec:single-pass-promptdsi-analysis}, we show that this design incurs only minor task performance degradation while speeding up both training and inference. We hypothesize that this is thanks to queries in document retrieval are semantically simple compared to the images in vision domain, leading to shallow query embeddings are sufficient for prompt selection.

\subsection{Topic-aware Prompt Keys}
\label{sec:topic-aware}
Given a prompt pool \(\mathbf{P}\), L2P~\cite{wang_l2p_cvpr22} shares \(\mathbf{P}\) for all incoming tasks, while S-Prompt++~\cite{wang_sprompt_neurips22,wang_hideprompt_neurips23} allocates a single key-prompt pair for each incoming tasks and freezes previous pairs.
Since the learnable prompt keys in the prompt pool are optimized to represent new corpora, the lack of distinct semantic boundaries among corpora in document retrieval (i.e, corpora both consist of documents of similar topics) causes these keys to become highly similar across corpora. This leads to a collapse in prompt selection (i.e., the query-key matching mechanism in Eq.~(\ref{eq:query_key_loss})) to a small subset of prompts, resulting in underutilization of parameters, which is visualized and elaborated in Section~\ref{subsec:prompt-selection-analysis}.

Instabilities in training prompt pools for PCL methods have been reported in the literature~\cite{moon_siblurry_iccv23,yadav_codecl_acl23}. 
We observe that in PCL methods, the query embeddings used for prompt selection are deterministic~\cite{wang_l2p_cvpr22}, suggesting that optimizing prompt keys \(\mathbf{K}\) might be unnecessary. Inspired by neural topic modeling techniques~\cite{gupta_topic_icml20,gerald_ircl_ecir22}, we propose using neural topic embeddings derived from \(D_0\) as fixed prompt keys. We employ BERTopic~\cite{grootendorst_bertopic_arxiv22} to cluster document embeddings and generate neural topic embeddings via a class-based TF-IDF procedure (we refer readers to~\cite{grootendorst_bertopic_arxiv22} for details of BERTopic).
By assuming each document semantically belongs to a neural topic, we employ these topic-aware fixed prompt keys to stabilize the query-key matching mechanism, addressing underutilization of prompts. Furthermore, this approach facilitates knowledge transfer between documents within the same topic and allows better interpretability compared to general-purpose or corpus-specific prompts in existing PCL methods. We refer to this PromptDSI variant as $\textbf{PromptDSI}_{\textbf{Topic}}$.

\subsection{Optimization Objective}
Denote PromptDSI's encoder and linear classifier as $\theta_e$ and $\theta_{l}$, respectively. At timestep $t=0$, PromptDSI is optimized using cross-entropy loss $\mathcal{L}_{\text{CE}}$ on $D_0$:
\begin{equation}
    \mathcal{L}_0 = \sum_{q\in Q_0} \mathcal{L}_{\text{CE}}(f_{\theta_l}(f_{\theta_e}(q)), \text{id}_q),
\end{equation} 
where $\text{id}_q \in I_0$ is the one-hot encoded ground truth docid of query $q$.

We study three popular PCL methods: L2P~\cite{wang_l2p_cvpr22}, S-Prompt++ (S++)~\cite{wang_sprompt_neurips22,wang_hideprompt_neurips23}, and CODA-Prompt (CODA)~\cite{smith_codaprompt_cvpr23}. 
During continual indexing, the encoder $\theta_{e}$ is frozen while prompt pool $\mathbf{P}$, prompt keys $\mathbf{K}$, and a portion of the linear classifier $\theta_{l}$ are optimized.
We refer to $f_{\mathbf{extra}, \theta_e}$ as PromptDSI's encoder parameterized by $\theta_e$ and a set of learnable components $\mathbf{extra}$. Denote $\mathbf{P}_t$ and $\mathbf{K}_t$ as prompts and prompt keys allocated for indexing $D_t$,  at timestep $t>0$, the general optimization objective for PromptDSI with L2P or S++ is:
\begin{equation}
    \label{eq:l2p_loss}
    \mathcal{L}_{t}^{\text{L2P}} = \sum\limits_{q\in Q_t} \min\limits_{\mathbf{P}_t, \mathbf{K}_t, \theta_{l,t}} \mathcal{L}_{\text{CE}}(f_{\theta_{l}}(f_{\mathbf{P}_t, \mathbf{K}_t, \theta_{e}}(q)), \text{id}_q) + \mathcal{L}_{\text{match}},
\end{equation}
where $\mathcal{L}_{\text{match}}$ is defined in Eq.~(\ref{eq:query_key_loss}) and $\text{id}_q \in I_t$. Unlike \(\textbf{PromptDSI}_{\textbf{L2P}}\), which optimizes all key-prompt pairs, \(\textbf{PromptDSI}_{\textbf{S++}}\) optimizes only one key-prompt pair per timestep. \(\textbf{PromptDSI}_{\textbf{CODA}}\) introduces learnable attention vectors \(A \in \mathbb{R}^{\text{dim}}\) for each prompt. Instead of using \(\mathcal{L}_{\text{match}}\), it computes a weighted sum of prompts:
\(P = \sum_{i=1}^{M} \alpha_i p_i,\)
where \(\alpha_i = \cos(h_q \odot A_i, k_i)\) and \(\odot\) denotes the Hadamard product. It is trained end-to-end with the following objective:
\begin{equation}
\label{eq:coda_loss}
\mathcal{L}_{t}^{\text{CODA}} = \sum\limits_{q\in Q_t} \min\limits_{\mathbf{P}_t, \mathbf{K}_t, A_t, \theta_{l,t}} \mathcal{L}_{\text{CE}}(f_{\theta_{l}}(f_{\mathbf{P}_t, \mathbf{K}_t, A_t, \theta_{e}}(q)), \text{id}_q).
\end{equation}
Using precomputed neural topic embeddings as fixed prompt keys, $\textbf{PromptDSI}_{\textbf{Topic}}$ omits optimizing prompt keys \(\mathbf{K}_t\) in Equation~\ref{eq:l2p_loss} and also removes $\mathcal{L}_{\text{match}}$:
\begin{equation}
\label{eq:topic_loss}
\mathcal{L}_{t}^{\text{Topic}} = \sum\limits_{q\in Q_t} \min\limits_{\mathbf{P}_t, \theta_{l,t}} \mathcal{L}_{\text{CE}}(f_{\theta_{l}}(f_{\mathbf{P}_t, \theta_{e}}(q)), \text{id}_q).
\end{equation}

%% file: overview-promptdsi.tex
\begin{figure*}[t]
    \centering
    \includegraphics[width=0.8\linewidth]{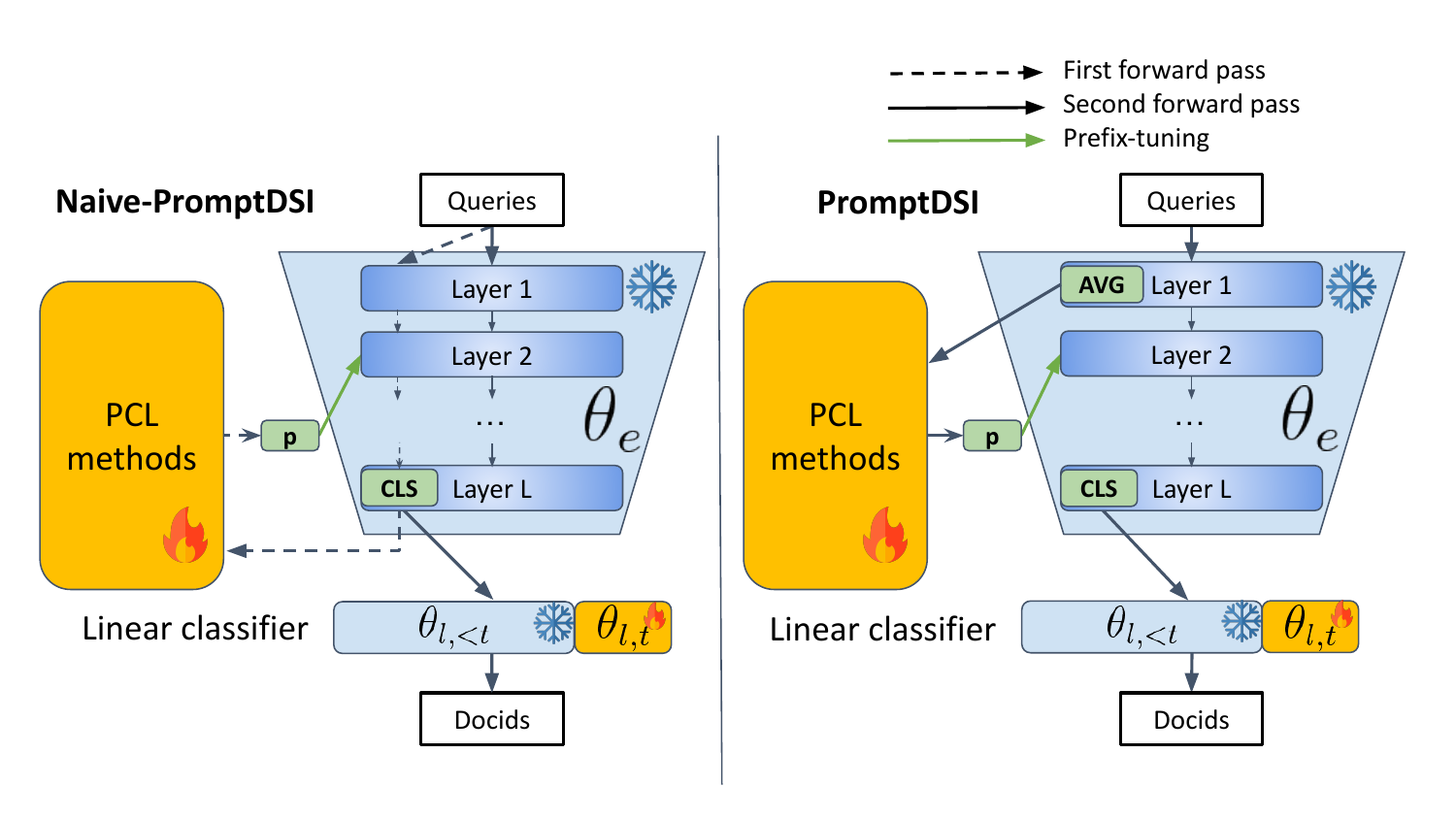}
    \caption{Naive-PromptDSI (left) integrates PCL methods into DSI using two forward passes, causing increased training/inference time. PromptDSI (right) enhances efficiency by using intermediate layer representations (i.e. average of token embeddings \texttt{[AVG]}) for prompt selection, effectively removing an additional forward pass. In this example, prompts \(p\) are prefix tuning to layer 2 of the DSI's encoder $\theta_e$. At timestep \(t>0\), $\theta_{l, <t}$ refers to the frozen portion of the linear classifier, while $\theta_{l, t}$ represents the expanded portion used for training on corpus $D_t$.}
    \label{fig:promptdsi-architecture}
\end{figure*}

%% file: 5_experiments.tex
\section{Experiments}

\subsection{Experimental Setting}

\subsubsection*{Datasets}
\label{subsec:datasets}
We evaluate PromptDSI on two well-known binary relevance document retrieval datasets: Natural Questions (NQ320k)~\cite{kwiatkowski_nq320k_tacl19} and a modified version of MSMARCO (MSMARCO 300k)~\cite{nguyen_msmarco_2016,craswell_msmarco_trec2019} (Table~\ref{tab:datasets}).
NQ320k refers to the title-de-duplicated version of the Natural Questions dataset, containing 320k query-document pairs from approximately 108k documents~\cite{wang_nci_neurips22,sun_genret_neurips23,kishore_incdsi_icml23}. 
MSMARCO 300k is a modified subset of the MS MARCO Document Ranking dataset~\cite{nguyen_msmarco_2016,craswell_msmarco_trec2019} with 300k documents, which is  established in previous works~\cite{mehta_dsipp_emnlp23,kishore_incdsi_icml23,sun_genret_neurips23}. For each corpus, queries from the official training set are split 80\%/20\% for training and validation, while the official development set is used for testing.
To mimic the continual learning (CL) setup, each dataset is split into an initial corpus ($D_0$) containing 90\% of the total documents, and five new corpora ($D_1$-$D_5$), each with 2\% of the total documents. Each document in the train set is supplied with up to 15 additional pseudo-queries generated using \texttt{docT5query}~\cite{nogueira_doct5query_2019} along with annotated natural queries. Each query corresponds to one relevant document.
\begin{table*}[t]
\small
\centering
\caption{The NQ320k and MSMARCO 300k dataset statistics used in our study.}
\label{tab:datasets}
\renewcommand{\arraystretch}{0.7} 
\scalebox{0.75}{
\begin{tabular}{cccccc}
\toprule
\multicolumn{6}{c}{NQ320k}\\ \midrule
\multicolumn{1}{c|}{Corpus} & \multicolumn{1}{c|}{Document} & \multicolumn{1}{c|}{Train Queries}  & \multicolumn{1}{c|}{Validation Queries}   & \multicolumn{1}{c|}{Test Queries} & Generated Queries \\ \midrule
\multicolumn{1}{c|}{$D_0$}     & \multicolumn{1}{c|}{98743}    & \multicolumn{1}{c|}{221194} & \multicolumn{1}{c|}{55295} & \multicolumn{1}{c|}{6998} & 1480538           \\
\multicolumn{1}{c|}{$D_1$}     & \multicolumn{1}{c|}{2000}     & \multicolumn{1}{c|}{4484}   & \multicolumn{1}{c|}{1091}  & \multicolumn{1}{c|}{152}  & 29997             \\
\multicolumn{1}{c|}{$D_2$}     & \multicolumn{1}{c|}{2000}     & \multicolumn{1}{c|}{4417}   & \multicolumn{1}{c|}{1085}  & \multicolumn{1}{c|}{153}  & 29992             \\
\multicolumn{1}{c|}{$D_3$}     & \multicolumn{1}{c|}{2000}     & \multicolumn{1}{c|}{4800}   & \multicolumn{1}{c|}{1298}  & \multicolumn{1}{c|}{177}  & 29991             \\
\multicolumn{1}{c|}{$D_4$}     & \multicolumn{1}{c|}{2000}     & \multicolumn{1}{c|}{4346}   & \multicolumn{1}{c|}{1107}  & \multicolumn{1}{c|}{116}  & 29992             \\
\multicolumn{1}{c|}{$D_5$}     & \multicolumn{1}{c|}{1874}     & \multicolumn{1}{c|}{4131}   & \multicolumn{1}{c|}{964}   & \multicolumn{1}{c|}{140}  & 28105 \\ 
\midrule
\multicolumn{6}{c}{MSMARCO 300k} \\ 
\midrule
\multicolumn{1}{c|}{$D_0$}     & \multicolumn{1}{c|}{289424}   & \multicolumn{1}{c|}{262008} & \multicolumn{1}{c|}{65502} & \multicolumn{1}{c|}{4678} & 4312150           \\
\multicolumn{1}{c|}{$D_1$}     & \multicolumn{1}{c|}{2000}     & \multicolumn{1}{c|}{1768}   & \multicolumn{1}{c|}{480}   & \multicolumn{1}{c|}{40}   & 29787             \\
\multicolumn{1}{c|}{$D_2$}     & \multicolumn{1}{c|}{2000}     & \multicolumn{1}{c|}{1799}   & \multicolumn{1}{c|}{457}   & \multicolumn{1}{c|}{35}   & 29805             \\
\multicolumn{1}{c|}{$D_3$}     & \multicolumn{1}{c|}{2000}     & \multicolumn{1}{c|}{1800}   & \multicolumn{1}{c|}{450}   & \multicolumn{1}{c|}{30}   & 29774             \\
\multicolumn{1}{c|}{$D_4$}     & \multicolumn{1}{c|}{2000}     & \multicolumn{1}{c|}{1772}   & \multicolumn{1}{c|}{475}   & \multicolumn{1}{c|}{29}   & 29821             \\
\multicolumn{1}{c|}{$D_5$}     & \multicolumn{1}{c|}{2000}     & \multicolumn{1}{c|}{1851}   & \multicolumn{1}{c|}{430}   & \multicolumn{1}{c|}{30}   & 29779 \\
\bottomrule
\end{tabular}
}
\end{table*}

\subsubsection*{Evaluation metrics.}
We adopt Hits{@}\{1, 10\} and Mean Reciprocal Rank (MRR){@}10 as document retrieval metrics.
To evaluate CL performance, after training on corpus $D_t$, we report average performance ($A_t$), forgetting ($F_t$), and learning performance ($LA_t$), following previous works \cite{mehta_dsipp_emnlp23}. 
In the main result tables, We report the results of initial corpus $D_0$ and new corpora $D_1$-$D_5$ separately. 
We emphasize on $A_t$ as it reflects both stability and plasticity~\cite{smith_codaprompt_cvpr23}. Let $P_{t,i}$ be the performance of the model on corpus $D_i$ in some metrics, after training on corpus $D_t$, where $i \leq t$. Forgetting of $D_0$ is calculated as: $\max(P_{5,0} - P_{0,0}, 0)$. With $t > 0$, the CL metrics are defined as follows:
\begin{equation*}
\centering
\begin{aligned}
A_t = \frac{1}{t}\sum_{i=1}^{t}P_{t,i} \quad LA_t = \frac{1}{t}\sum_{i=1}^{t}P_{i,i} \quad 
F_t = \frac{1}{t}\sum_{i=0}^{t-1}\max_{i^{'}\in\{0,\ldots,t-1\}} (P_{i^{'},i} - P_{t,i})
\end{aligned}
\end{equation*}

\subsubsection*{Baselines.}
We adopt the well-established BERT~\cite{devlin_bert_naacl19} and SBERT~\cite{reimers_sbert_emnlp19}\footnote{HuggingFace model identifiers: \texttt{google-bert/bert-base-uncased} and \texttt{sentence-transformers/all-mpnet-v2}} as backbones and compare PromptDSI with both CL and non-CL baselines.\footnote{We omit comparisons with regularization-based methods, as they underperform compared to replay-based approaches~\cite{wu_protoprobing_iclr21}.} 
While recent dense encoders such as BGE~\cite{xiao_bge_sigir24}, and NV-Embed~\cite{lee_nvembed_iclr25} can also serve as backbones and may yield improved retrieval performance, our goal is not to benchmark classification-based DSI frameworks with state-of-the-art backbones, but to analyze their continual learning behavior. Due to resource and space constraints, we leave such comparisons to future work.

\begin{itemize}
    \item \textbf{Sequential Fine-tuning} sequentially optimizes on new corpora without accessing previous ones, serving as the performance lower bound in CL.
      \item \textbf{DSI++~\cite{mehta_dsipp_emnlp23}} involves sequential fine-tuning on new corpora, using a query generation model to generate pseudo queries for sparse experience replay.
    \item \textbf{IncDSI~\cite{kishore_incdsi_icml23} } indexes new documents by sequentially caching previous query embeddings and expanding $\theta_l$. 
    The new $\theta_l$'s embeddings are determined by solving constrained optimization problems. 
    Since IncDSI is designed for online learning, we include ``IncDSI*'', a variant that solves the optimization for several epochs.
    We regard IncDSI as a strong baseline on $D_0$.
    \item \textbf{Multi-corpora Fine-tuning (Multi)} only fine-tunes on new corpora $D_1$-$D_5$, which are merged as one corpus. 
    It is equivalent to the multi-task learning baseline in CL, serving as the performance upper bound on new corpora.
    \item \textbf{Joint Supervised (Joint)} trains on both initial and all new corpora, similar to the conventional supervised learning setup.
    \item \textbf{Dense Passage Retrieval (DPR)~\cite{karpukhin_dpr_emnlp20}} is a popular BERT-based dual-encoder trained with BM25~\cite{robertson_bm25_2009} hard negatives and in-batch negatives. DPR serves as a strong dense retrieval baseline and performs zero-shot retrieval on new corpora.
\end{itemize}

\subsubsection*{Implementation Details.}
We employ AdamW~\cite{loshchilov_adamw_arxiv17} and use a single NVIDIA A100 80GB GPU for all experiments.   
Following~\cite{kishore_incdsi_icml23}, we train BERT/SBERT on \(D_0\) for 20 epochs, using batch size 128 and 1024, learning rate \(1e^{-4}\) and \(5e^{-5}\) for NQ320k and MSMARCO 300k, respectively. All subsequent methods are trained with batch size 128 and initialized from the same BERT/SBERT checkpoint trained on \(D_0\).
We randomly sample pseudo-queries from previous corpora to substitute for the query generation model in DSI++\footnote{Since the code for DSI++ has not been released.}.
We reproduce IncDSI~\cite{kishore_incdsi_icml23} using its official implementation.
For PCL methods in PromptDSI, we leverage open-source implementations. \(\text{PromptDSI}_{\text{L2P/S++}}\) use a prompt pool of size 5, prompt length 20 and top-1 prompt selection. \(\text{PromptDSI}_{\text{CODA}}\) uses a prompt length 10 and 2 prompts per task. For BERT-based PromptDSI, we use learning rate \(1e^{-4}\) and \(5e^{-4}\); for SBERT-based PromptDSI, we use \(1.5e^{-4}\) and \(1e^{-3}\) for NQ320k and MSMARCO 300k, respectively. For \(\text{PromptDSI}_{\text{Topic}}\), we use BERTopic~\cite{grootendorst_bertopic_arxiv22} to mine neural topics from \(D_0\), adhering closely to the author's best practices guide
\footnote{\url{https://maartengr.github.io/BERTopic/getting\_started/best\_practices/\\best\_practices.html}}.
PromptDSI variants are trained for 10 epochs since prefix-tuning~\cite{li_prefixtuning_21} requires longer training to converge. Other full-model fine-tuning baselines are trained for 5 epochs. Results for CL methods are averaged over three runs, with standard deviations reported accordingly. The layer-wise prompting study in Section~\ref{subsec:layerwise-prompting-study} is conducted using a fixed random seed.

\input{main-result-all-simplified-bert}
\input{continual_indexing_bert}
\input{main-result-all-simplified-sbert}

\subsection{Main Results}
\label{subsec:results}
We present our results of BERT/SBERT-based methods in Table~\ref{tab:bert_main_result} and~\ref{tab:sbert_main_result}. Performance of the latter is often better thanks to the better representation.

Among non-CL methods, across backbones and datasets, \textbf{Multi-corpora Fine-tuning} completely forgets \(D_0\) and heavily overfits \(D_1\)-\(D_5\).
Both \textbf{Joint} and \textbf{DPR} achieve a balance between initial and new corpora; however, they generally underperform IncDSI and PromptDSI.
Among CL methods, \textbf{Sequential Fine-tuning} suffer sfrom severe catastrophic forgetting and significantly overfitting the most recent corpora.
\textbf{DSI++}, even with sparse experience replay, provides only slight improvements over Sequential Fine-tuning, highlighting that maintaining a memory buffer or a generative model for rehearsal is non-trivial.
Overall, Sequential Fine-tuning and DSI++ employ full-model fine-tuning CL methods (i.e., huge trainable parameters), but suffers from catastrophic forgetting. 
In contrast, IncDSI and PromptDSI are significantly more parameter-efficient, as a result of freezing the backbone.

\textbf{IncDSI} maintains strong performance across all corpora, with IncDSI* (10 epochs) further improving overall metrics. However, it still falls short of optimal performance on new corpora. All \textbf{PromptDSI} variants outperform IncDSI and IncDSI* in terms of $A_5$ by large margins across metrics, datasets, and backbones while maintaining \(D_0\) performance close to that of IncDSI. \textbf{PromptDSI\(_{\text{Topic}}\)} removes prompt-key optimization to stabilize query-key matching, achieving comparable or superior performance among PromptDSI variants.  

We further analyze the continual indexing of BERT-based PromptDSI and IncDSI in Figures~\ref{fig:continual_indexing_nq320k} and~\ref{fig:continual_indexing_msmarco}. PromptDSI consistently outperforms IncDSI in Average and Learning Performance across datasets, with minimal forgetting, often matching IncDSI or surpassing IncDSI*. These results highlight the superior stability-plasticity trade-off of PromptDSI, despite being rehearsal-free.  

\subsection{Memory Complexity Analysis}
DSI++ requires storing an additional T5~\cite{raffel_t5_jmlr20} model for query generation and IncDSI requires caching a matrix $\mathbf{Z} \in \mathbb{R}^{\text{dim}\times\sum_{t=0}^{T}|D_t|}$, where each column of $\mathbf{Z}$ represents an average query embedding.
Consequently, the memory usage and indexing time of IncDSI increase linearly with the number of documents. For our experiments, caching requires approximately 318\,MiB for NQ320k and 977\,MiB for MSMARCO 300k. 
Extending to the full MS MARCO dataset (8.8M passages) requires about 25\,GiB of memory. In many cases, loading such a large memory footprint onto conventional GPUs may not be possible. In contrast, PromptDSI is rehearsal-free, eliminating the need to store \(\mathbf{Z}\). It employs a prompt pool and a set of prompt keys, i.e., \(\textbf{P}\cup \textbf{K} \in \mathbb{R}^{\text{dim}\times (M(m+1))}\), which consists of \(M\) prompts \(p \in \mathbb{R}^{\text{dim}\times m}\) and \(M\) prompt keys \(k \in \mathbb{R}^{\text{dim}}\). The L2P/S++ variants require approximately 315\,KiB, which is about three orders of magnitude smaller than IncDSI's overhead on NQ320k, while the largest “Topic” variant requires approximately 11.9\,MiB—roughly one to two orders of magnitude smaller than IncDSI or the additional T5 model used in DSI++.

\subsection{Single-pass PromptDSI Analysis}
\label{subsec:single-pass-promptdsi-analysis}

Table~\ref{tab:ospromptdsi} presents the comparison between Naive-PromptDSI and PromptDSI. Using intermediate layer representations as query embeddings for prompt selection (i.e., bypassing the first forward pass), we observe only a minimal trade-off in performance across all PromptDSI variants. This is primarily reflected in MRR@10, with a maximum drop of just 0.9\%. Notably, with our implementations, the single-pass L2P and S++ variants can be up to 4 times faster by omitting the first forward pass. Although the speedup is less significant for the CODA variant, which employs a weighted sum of prompts strategy, it still achieves a notable 1.7x speedup. Overall, these results confirm that the proposed single-pass PCL methods in PromptDSI meet the low-latency requirements of typical information retrieval systems, with only a negligible impact on performance.

\input{naive-vs-promptdsi}
\input{fig-selection-percentage}
\input{prompt_topic_examples}

\subsection{Prompt Pool Utilization Analysis}
\label{subsec:prompt-selection-analysis}
Figure~\ref{fig:prompt-selection} (left) illustrates the prompt pool utilization of PromptDSI variants.\footnote{PromptDSI$_\text{CODA}$ applies a weighted sum of prompts instead of prompt selection.}. We observe that L2P frequently selects a small subset of prompts across all corpora. While S++ benefits from SBERT’s better similarity-based embeddings to achieve more diverse task-specific prompt selection, this effect does not extend to L2P. These findings suggest that optimizing the prompt pool and ensuring diverse selection remain challenging due to instability.
With a topic-aware prompt pool, PromptDSI$_\text{Topic}$ achieves better prompt utilization, leading to improved performance in some cases (Table~\ref{tab:sbert_main_result}) while mitigating training instability. Notably, in Figure~\ref{fig:prompt-selection} (bottom-right), prompt pool utilization exceeds 60\%, with over 30\% of the prompts frequently selected (i.e., above the uniform threshold). Beyond performance improvements, topic-aware prompts enhance interpretability, unlike general-purpose (L2P) or corpus-specific (S++) prompts.

\begin{figure*}[ht]
      \centering
      \includegraphics[width=0.45\linewidth]{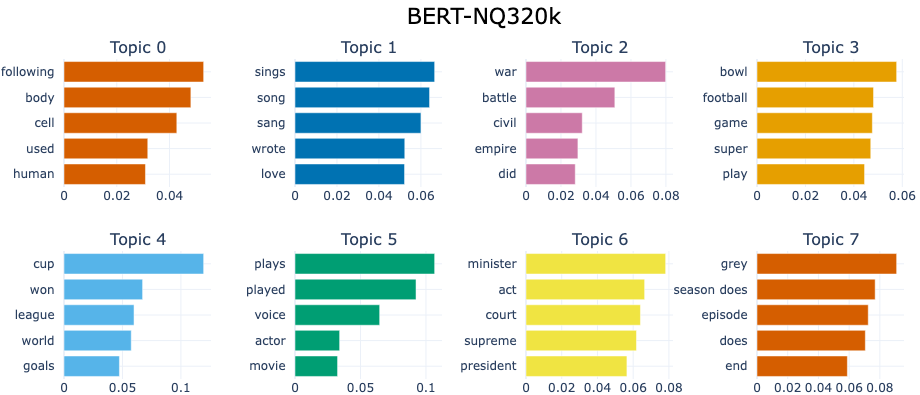} \rulesep
      \includegraphics[width=0.45\linewidth]{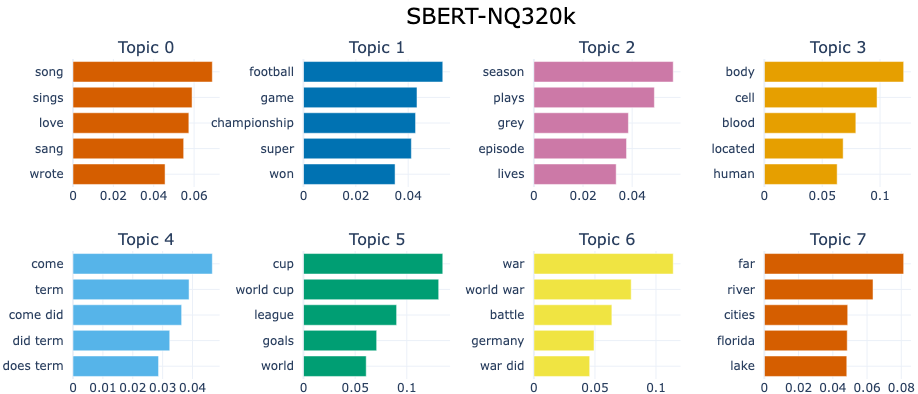}
      \caption{Top 8 NQ320k topics mined using BERTopic with BERT/SBERT and their corresponding most frequent terms. 
      A total of 80 topics were identified with BERT and 91 with SBERT from $D_0$.
      }
      \label{fig:nq320k-topics}
\end{figure*}

\begin{figure*}[ht]
      \centering
      \includegraphics[width=0.49\linewidth]{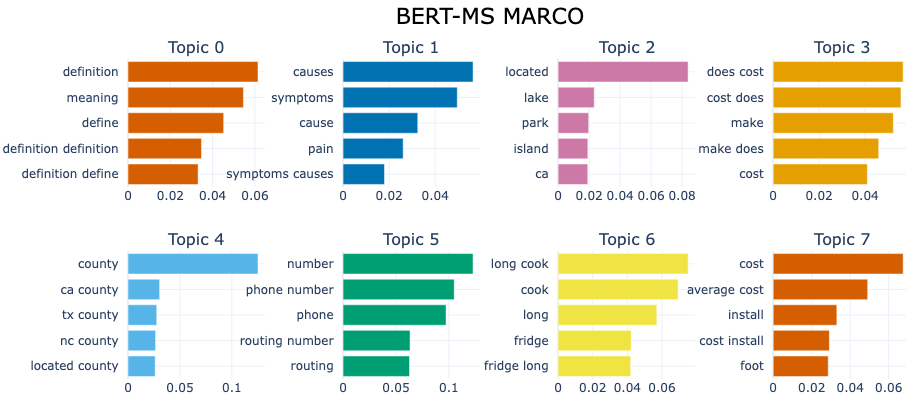} \rulesep
      \includegraphics[width=0.49\linewidth]{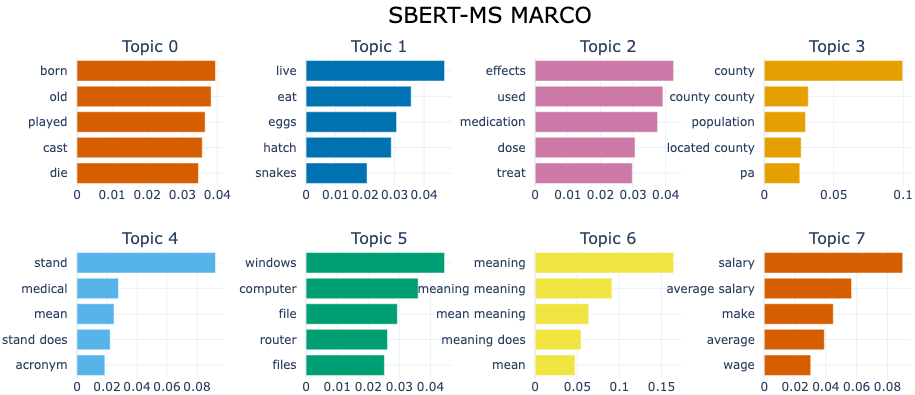}
      \caption{Top 8 MSMARCO 300k topics mined using BERTopic with BERT/SBERT and their corresponding most frequent terms. 
      A total of 182 topics were identified with BERT and 193 with SBERT from $D_0$. 
      }
      \label{fig:msmarco-topics}
\end{figure*}

Since we uses two backbones BERT and SBERT, we adopt separate BERTopic models for each, resulting in slightly different topics for the same dataset (Figures~\ref{fig:nq320k-topics} and~\ref{fig:msmarco-topics}). While effective, topic modeling may benefit further from LLMs due to their superior representational and contextual capabilities.
Table~\ref{tab:topics-queries} presents queries associated with the top 8 topics identified by SBERT-based BERTopic on the NQ320k validation set, which PromptDSI$_\text{Topic}$ selects during inference. These neural topics reflect frequently queried subjects, with queries often containing recurring or semantically similar terms. 

\input{fig-prompting-analysis}
\subsection{Layer-wise Prompting Study}
\label{subsec:layerwise-prompting-study}
To determine the optimal layers for prompting (i.e., prefix-tuning), we follow the protocol of~\cite{wang_dualprompt_eccv22}. We assume prompting layers are contiguous and limit the search space to three layers. Using SBERT-based PromptDSI, we attach prompts to each layer of $\theta_e$ to identify effective layers for multi-layer prompting. Given that PromptDSI variants improve $A_5$ at the cost of a slight degradation in $D_0$, we define our main criteria as follows: (1) \textit{Lower forgetting on $D_0$}, (2) \textit{Moderate performance on $A_5$}, and (3) \textit{Fewer prompting layers are preferable}. For single-layer prompting, Figure~\ref{fig:sbert_single_layer_study} (left) shows that CODA retains more knowledge but performs worse than L2P and S++. Overall, PromptDSI outperforms IncDSI across prompting layers. However, L2P and S++ exhibit severe forgetting in the top layers (10-12), leading us to exclude them from further analysis. Notably, layer 2 achieves comparable or lower forgetting than IncDSI on $D_0$ while maintaining strong $A_5$, making it a promising choice. We select S++ for multi-layer prompting analysis due to its high forgetting on $D_0$. As shown in Figure~\ref{fig:sbert_single_layer_study} (right), two-layer prompting (first column) slightly reduces forgetting, but the gains in $A_5$ are marginal ($<1\%$ for MRR@10, $<0.5\%$ for Hits@10). Three-layer prompting significantly degrades MRR@10 (up to $2.5\%$) despite increased computation, offering minimal reduction in forgetting. Since multi-layer prompting incurs higher computational costs, these findings suggest that single-layer prompting, particularly in the lower layers (1-5), achieves the best trade-off between efficiency and performance. Consequently, we primarily perform single-layer prefix-tuning on layer 2 of DSI’s encoder in our experiments, which satisfies all desirable criteria.

%% file: main-result-all-simplified-bert.tex
\begin{table*}[t]
\caption{
BERT-based methods performance after indexing \(D_5\). 
H@10 and M@10 denote Hits@10 and MRR@10.
\(\mathbf{D_0}\) denotes \(P_{5,0}\) on \(D_0\)'s test queries.
$\dagger$ denotes results from~\cite{kishore_incdsi_icml23}.
Params. denotes number of trainable parameters.
\textbf{Bold} and \underline{underline} highlight the top and second best CL methods.
The underscript numbers denote the standard deviations.
}
\label{tab:bert_main_result}
\centering
\renewcommand{\arraystretch}{0.7} 
\scalebox{0.75}{
\begin{tabular}{l|ccccc|ccccc|c}

\toprule
{} & \multicolumn{5}{c|}{\textbf{NQ320k}} & \multicolumn{5}{c|}{\textbf{MSMARCO 300k}} \\
\midrule

\textbf{BERT-} & \multicolumn{2}{c}{$\mathbf{D_0}\uparrow$} & \multicolumn{2}{c}{$A_5\uparrow$} & \multirow{2}{*}{Params.$\downarrow$} & \multicolumn{2}{c}{$\mathbf{D_0\uparrow}$} & \multicolumn{2}{c}{$A_5\uparrow$} & \multirow{2}{*}{Params.$\downarrow$} & {Rehearsal} \\ 

\textbf{based} & H@10 & M@10 & H@10 & M@10 & & H@10 & M@10 & H@10 & M@10 & & {free}\\
\midrule

\multicolumn{9}{l}{\textbf{Non CL Methods (For Reference)}} \\
Multi & 0.0 & 0.0 & 91.4 & 83.9 & 193\,M & 0.0 & 0.0 & 92.2 & 83.9 & 340\,M & -\\

Joint & 85.9 & 70.1 & 84.7 & 68.4 & 193\,M & 76.8 & 55.2 & 78.3 & 53.8 & 340\,M & -\\

DPR & 70.3 & 51.9 & 70.1 & 49.8 & 220\,M & 68.8$^\dagger$ & - & 62.8$^\dagger$ & - & 220\,M & -\\
\midrule

\multicolumn{9}{l}{\textbf{CL Methods}} \\  
Sequential & 
$0.0_{0.0}$ & $0.0_{0.0}$ & $27.4_{2.8}$ & $22.2_{1.2}$ & 193\,M & $0.0_{0.0}$ & $0.0_{0.0}$ & $31.9_{0.8}$ & $27.2_{0.6}$  & 340\,M & \cmark \\

DSI++ & $2.6_{0.0}$ & $2.6_{2.6}$ & $28.6_{1.9}$ & $22.1_{28.6}$ & 193\,M & $2.6_{0.1}$ & $2.4_{0.0}$ & $28.6_{7.3}$ & $24.2_{5.1}$ & 340\,M & \xmark \\

IncDSI & $\underline{86.4}_{0.1}$ & $\textbf{72.2}_{0.3}$ & $85.8_{0.6}$ & $69.5_{0.3}$ & \textbf{7.6\,M} & $79.5_{0.9}$ & $57.5_{1.3}$ & $81.8_{2.2}$ & $61.4_{3.2}$ & \textbf{7.7\,M} & \xmark \\ 

IncDSI* & $\textbf{86.5}_{0.2}$ & $\underline{72.0}_{0.1}$ & $85.2_{1.0}$ & $68.6_{0.9}$ & \textbf{7.6\,M} & $\textbf{80.6}_{0.0}$ & $\textbf{59.0}_{0.0}$ & $84.8_{0.1}$ & $65.1_{0.0}$ & \textbf{7.7\,M} & \xmark\\

\midrule

\multicolumn{9}{l}{\textbf{PromptDSI (Ours) with}} \\  
        
$\textbf{L2P}$ & $86.1_{0.1}$ & $71.2_{0.2}$ & $\textbf{90.8}_{0.4}$ & $73.2_{1.8}$ & \textbf{7.6\,M} & $\underline{80.5}_{0.0}$ & $58.7_{0.0}$ & $86.7_{1.4}$ & $\underline{67.4}_{1.1}$ & \underline{7.8\,M} & \cmark \\

\textbf{S++} & $86.1_{0.2}$ & $71.4_{0.1}$ & $\underline{90.5}_{0.4}$ & $\underline{73.8}_{0.7}$ & \textbf{7.6\,M} & $80.4_{0.1}$ & $58.6_{0.1}$ & $\underline{86.8}_{1.5}$ & $\underline{67.4}_{0.8}$ & \underline{7.8\,M} & \cmark \\

\textbf{CODA} & $86.0_{0.0}$ & $71.2_{0.1}$ & $90.3_{0.3}$ & $\textbf{74.2}_{0.8}$ & \underline{7.7\,M} & $\textbf{80.6}_{0.1}$ & $\underline{58.9}_{0.2}$ & $\textbf{87.9}_{0.6}$ & $66.8_{0.3}$ & \underline{7.8\,M} & \cmark \\

\textbf{Topic} & $86.0_{0.1}$ & $71.3_{0.0}$ & $\underline{90.5}_{0.5}$ & $\textbf{74.2}_{1.4}$ & 8.8\,M & $80.4_{0.0}$ & $58.6_{0.1}$ & $86.7_{0.5}$ & $\textbf{67.5}_{0.5}$ & 10.4\,M & \cmark \\
\bottomrule
\end{tabular}
}
\end{table*}

%% file: continual_indexing_bert.tex
\begin{figure}[t]
      \centering      
      \includegraphics[width=0.72\linewidth]{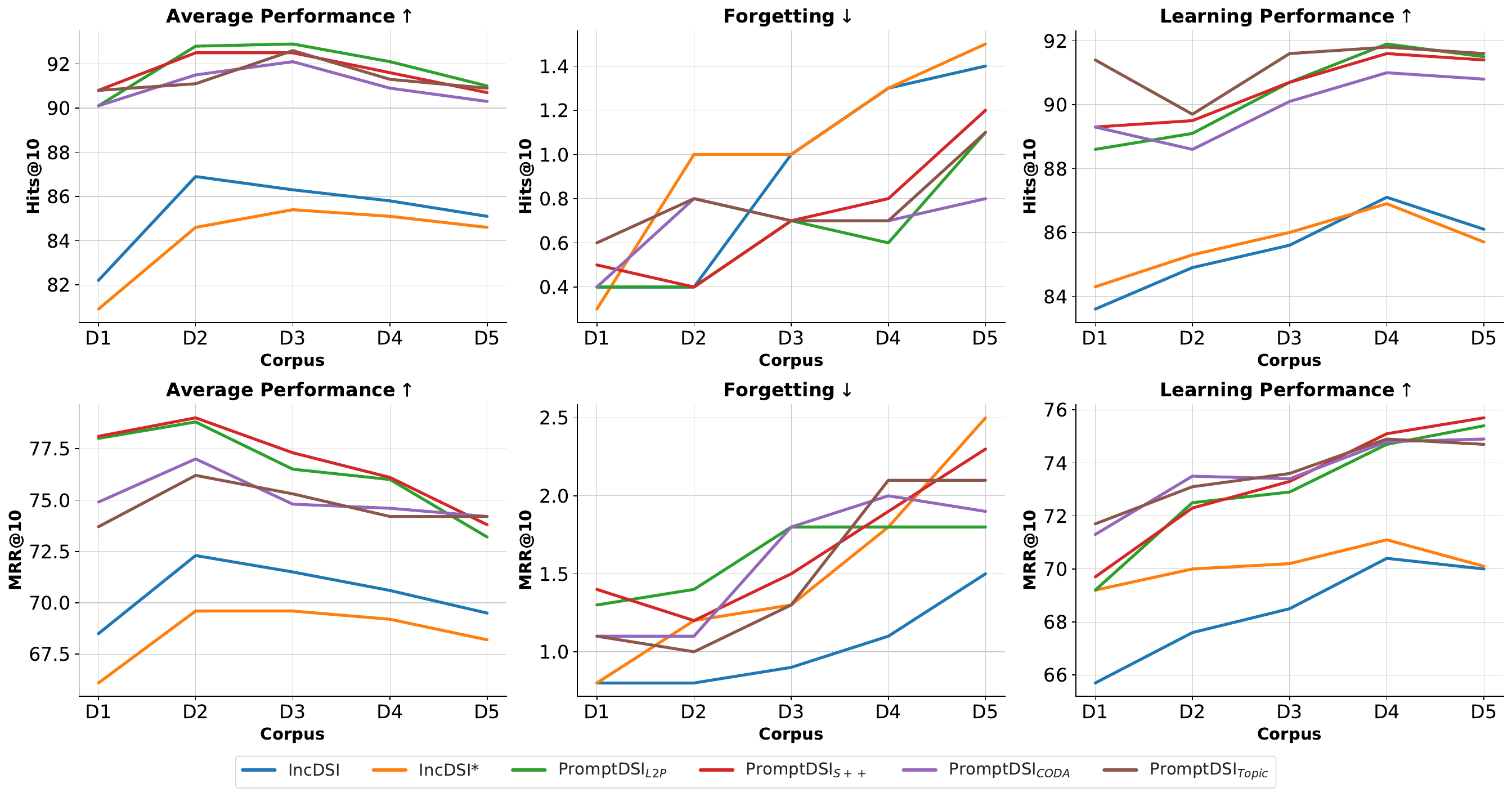}
      \caption{Continual indexing performance of BERT-based methods on NQ320k}
      \label{fig:continual_indexing_nq320k}
\end{figure}

\begin{figure}[t]
      \centering      
      \includegraphics[width=0.72\linewidth]{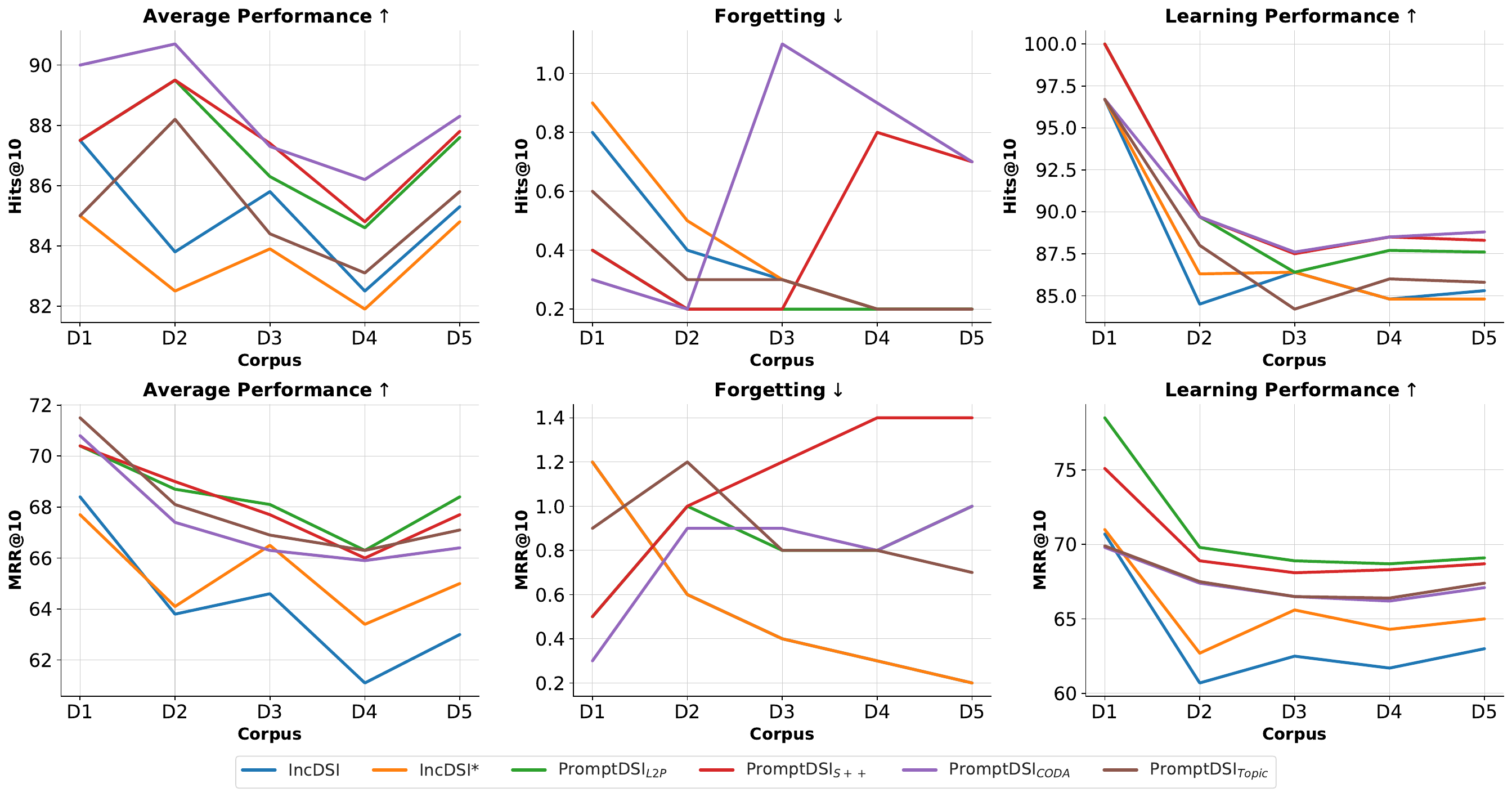}
      \caption{Continual indexing performance of BERT-based methods on MSMARCO 300k}
      \label{fig:continual_indexing_msmarco}
\end{figure}

%% file: main-result-all-simplified-sbert.tex
\begin{table*}[t]
\caption{
SBERT-based methods performance after indexing \(D_5\). 
\(\mathbf{D_0}\) denote \(P_{5,0}\) on \(D_0\)'s test queries.
H@10 and M@10 denotes Hits@10 and MRR@10.
Params. denotes number of trainable parameters.
\textbf{Bold} and \underline{underline} highlight the top and second best CL methods. The underscript numbers denote the standard deviations.
}
\label{tab:sbert_main_result}
\centering
\renewcommand{\arraystretch}{0.7} 
\scalebox{0.75}{
\begin{tabular}{l|ccccc|ccccc|c}

\toprule
{} & \multicolumn{5}{c|}{\textbf{NQ320k}} & \multicolumn{5}{c|}{\textbf{MSMARCO 300k}} \\
\midrule

\textbf{SBERT-} & \multicolumn{2}{c}{$\mathbf{D_0}\uparrow$} & \multicolumn{2}{c}{$A_5\uparrow$} & \multirow{2}{*}{Params.$\downarrow$} & \multicolumn{2}{c}{$\mathbf{D_0\uparrow}$} & \multicolumn{2}{c}{$A_5\uparrow$} & \multirow{2}{*}{Params.$\downarrow$} & {Rehearsal} \\ 

\textbf{based} & H@10 & M@10 & H@10 & M@10 & & H@10 & M@10 & H@10 & M@10 & & {free}\\
\midrule

\multicolumn{9}{l}{\textbf{Non CL methods (For reference)}} \\
Multi & 0.0 & 0.0 & 92.0 & 83.6 & 193\,M & 0.0 & 0.0 & 94.7 & 87.2 & 340\,M & - \\

Joint & 86.8 & 71.5 & 86.4 & 69.7 & 193\,M & 76.3 & 55.5 & 84.1 & 61.2 & 340\,M & - \\

\midrule

\multicolumn{9}{l}{\textbf{CL methods}} \\  
Sequential & $0.0_{0.0}$ & $0.0_{0.0}$ & $25.2_{2.0}$ & $20.3_{0.8}$ & 193\,M & $0.0_{0.0}$ & $0.0_{0.0}$ & $27.1_{2.1}$ & $23.3_{1.2}$ & 340\,M & \cmark \\

DSI++ & $2.7_{0.0}$ & $2.4_{0.1}$ & $28.5_{4.8}$ & $22.3_{2.7}$ & 193\,M & $2.8_{0.1}$ & $2.5_{0.1}$ & $29.0_{4.7}$ & $23.3_{0.5}$ & 340\,M & \xmark \\

IncDSI & $\textbf{87.1}_{0.0}$ & $72.3_{1.1}$ & $86.6_{0.8}$ & $70.5_{2.4}$ & \textbf{7.6\,M} & $80.9_{1.0}$ & $58.6_{1.2}$ & $82.8_{1.6}$ & $64.0_{3.3}$ & \textbf{7.7\,M} & \xmark \\

IncDSI* & $87.0_{0.0}$ & $\textbf{72.6}_{0.0}$ & $87.3_{0.1}$ & $73.2_{0.0}$ & \textbf{7.6\,M} & $\textbf{82.0}_{0.0}$ & $\textbf{60.0}_{0.0}$ & $84.1_{0.1}$ & $68.0_{0.2}$ & \textbf{7.7\,M} & \xmark \\

\midrule

\multicolumn{9}{l}{\textbf{PromptDSI (Ours) with}} \\  
$\textbf{L2P}$ & $86.9_{0.1}$ & $\textbf{72.6}_{0.1}$ & $\textbf{91.1}_{0.1}$ & $74.0_{3.8}$ & \textbf{7.6\,M} & $\underline{81.6}_{0.1}$ & $\underline{59.1}_{0.0}$ & $87.4_{0.4}$ & $70.1_{0.4}$ & \underline{7.8\,M} & \cmark \\

\textbf{S++} & $86.8_{0.0}$ & $\underline{72.5}_{0.1}$ & $\underline{91.0}_{0.5}$ & $\underline{74.8}_{2.7}$ & \textbf{7.6\,M} & $81.4_{0.1}$ & $58.9_{0.1}$ & $\underline{87.5}_{0.9}$ & $\textbf{71.4}_{0.3}$ & \underline{7.8\,M} & \cmark \\

\textbf{CODA} & $\underline{87.0}_{0.1}$ & $72.1_{0.9}$ & $\textbf{91.1}_{0.5}$ & $74.2_{4.1}$ & \underline{7.7\,M} & $81.5_{0.0}$ & $59.0_{0.2}$ & $\underline{87.5}_{0.7}$ & $70.0_{1.0}$ & \underline{7.8\,M} & \cmark \\

\textbf{Topic} & $86.8_{0.0}$ & $72.1_{0.2}$ & $\textbf{91.1}_{0.3}$ & $\textbf{75.1}_{1.9}$ & 9.0\,M & $81.3_{0.1}$ & $\underline{59.1}_{0.2}$ & $\textbf{88.1}_{1.9}$ & $\underline{70.5}_{1.8}$ & 10.6\,M & \cmark \\

\bottomrule

\end{tabular}
}
\end{table*}

%% file: naive-vs-promptdsi.tex
\begin{table}[t]
\caption{Performance comparison between Naive-PromptDSI and PromptDSI on NQ320k.}
\label{tab:ospromptdsi}
\centering
\renewcommand{\arraystretch}{0.7} 
\scalebox{0.75}{
\begin{tabular}{l|c|cc|cc|c}
\toprule
\multirow{2}{*}{Methods} & {\multirow{2}{*}{Metric}} & \multicolumn{2}{c|}{Naive-PromptDSI}  & \multicolumn{2}{c|}{PromptDSI} & \multirow{2}{*}{Single-pass speedup} \\ 
{} & {} & Hits@10 & MRR@10 &  Hits@10 &  MRR@10 \\  \midrule

\multirow{2}{*}{PromptDSI$_\text{L2P}$}
 & \(\mathbf{D_0}\uparrow\) & $\text{86.1}_{0.1}$ & $\text{71.3}_{0.1}$ & $\text{86.1}_{0.1}$ & $\text{71.2}_{0.2}$ & \multirow{2}{*}{4.3x} \\
 & $A_5\uparrow$ & $\text{90.3}_{0.5}$ & $\text{74.1}_{0.3}$ & $\text{90.8}_{0.4}$ & $\text{73.2}_{1.8}$ \\
\midrule
\multirow{2}{*}{PromptDSI$_\text{S++}$}
 & \(\mathbf{D_0}\uparrow\) & $\text{86.1}_{0.1}$ & $\text{71.3}_{0.2}$ & $\text{86.1}_{0.2}$ & $\text{71.4}_{0.1}$ & \multirow{2}{*}{4.2x}\\
 & $A_5\uparrow$ & $\text{90.5}_{0.2}$ & $\text{74.0}_{0.9}$ & $\text{90.5}_{0.4}$ & $\text{73.8}_{0.7}$ \\
\midrule
\multirow{2}{*}{PromptDSI$_\text{CODA}$}
 & \(\mathbf{D_0}\uparrow\) & $\text{86.0}_{0.1}$ & $\text{71.2}_{0.1}$ & $\text{86.0}_{0.0}$ & $\text{71.2}_{0.1}$ & \multirow{2}{*}{1.7x}\\
 & $A_5\uparrow$ & $\text{90.6}_{0.2}$ & $\text{74.2}_{0.6}$ & $\text{90.3}_{0.3}$ & $\text{74.2}_{0.8}$ \\
\bottomrule
\end{tabular}
}
\end{table}

%% file: fig-selection-percentage.tex
\begin{figure}[t]
      \centering      
      \includegraphics[width=0.55\linewidth]{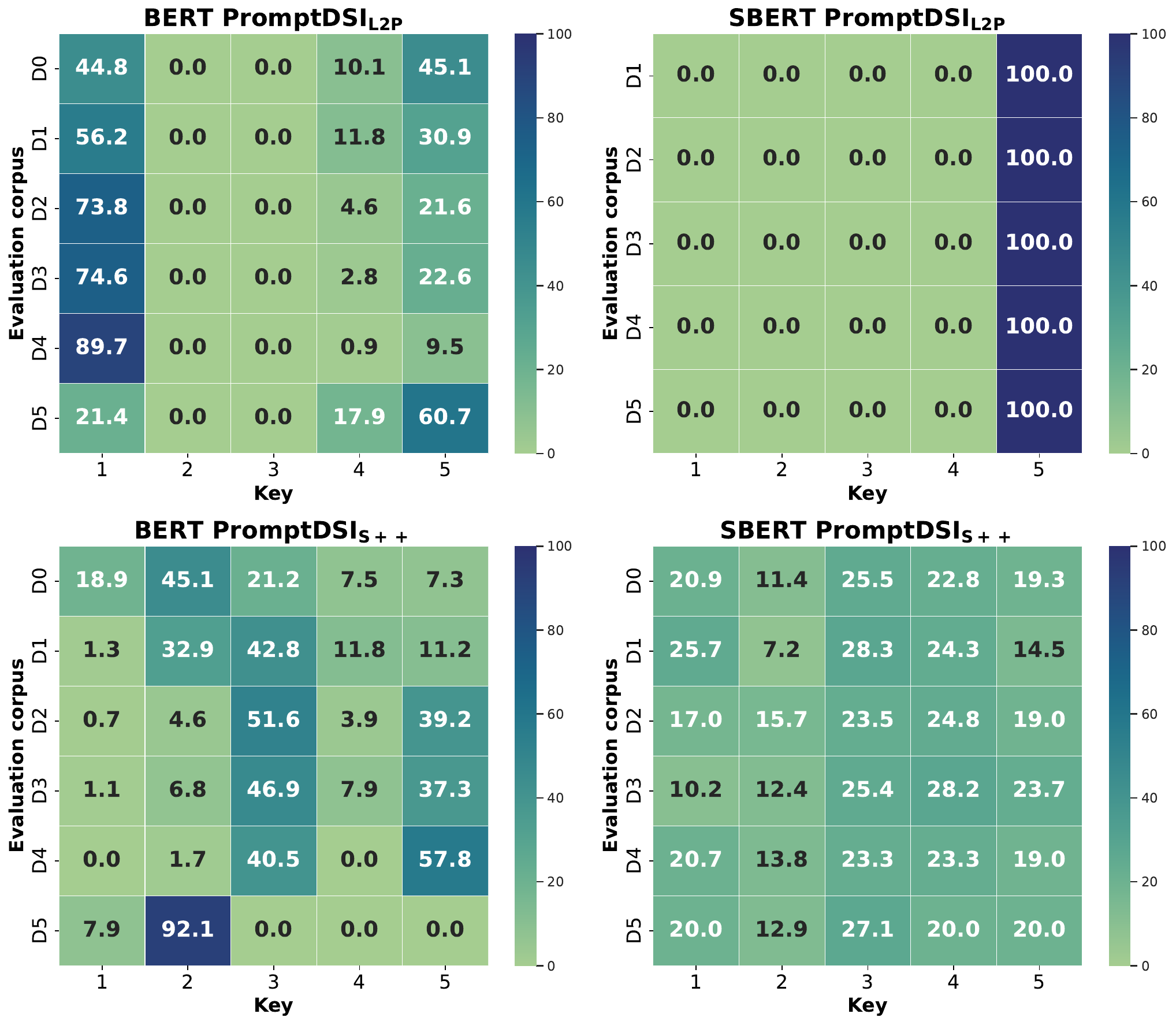}
      \rulesep
      \includegraphics[width=0.25\linewidth]{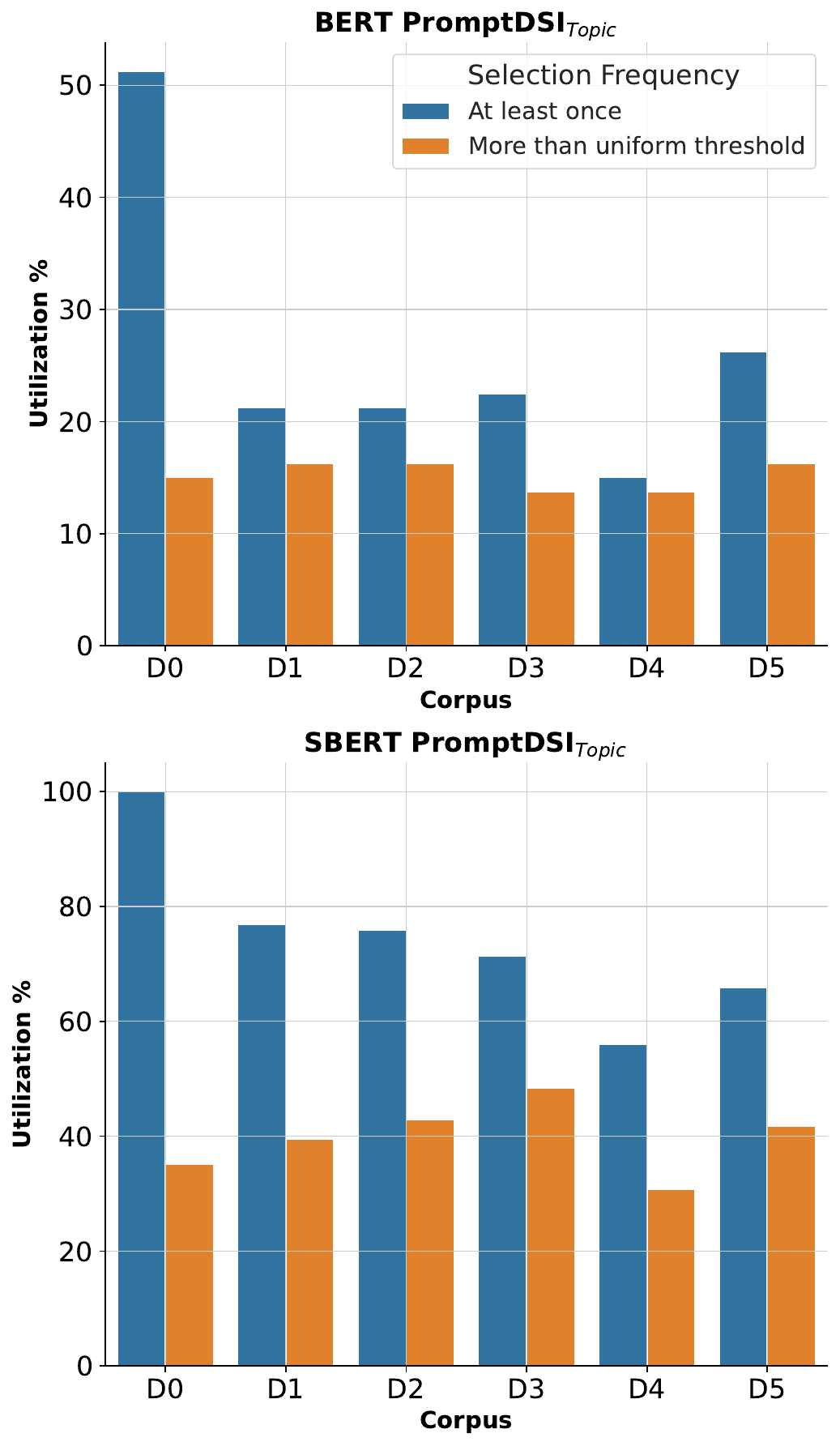}
      \caption{Prompt pool utilization:PromptDSI$_\text{L2P/S++}$ (left); PromptDSI$_\text{Topic}$ (right).}
      \label{fig:prompt-selection}
\end{figure}

%% file: prompt_topic_examples.tex
\begin{table*}[t]
\caption{Examples annotated queries from the NQ320k validation set that are matched correctly to the corresponding topic-aware prompts during inference. Topic embeddings are mined using SBERT-based BERTopic. Topic-specific terms are highlighted.}
\label{tab:topics-queries}
\small
\centering
\renewcommand{\arraystretch}{0.5}
\scalebox{0.7}{
\begin{tabular}{l|l}
\toprule
\multicolumn{1}{c|}{\textbf{Topic}} & \multicolumn{1}{c}{\textbf{Queries}} \\ \midrule
0 (Music)     & \begin{tabular}[c]{@{}l@{}}
who \textbf{sang} the \textbf{song} no other love have i\\ 
who \textbf{sings} the \textbf{song} it feels like rain\\ 
who \textbf{sings} whenever you want me i'll be there\\ 
\end{tabular} \\ 
\midrule
1 (Football)     & \begin{tabular}[c]{@{}l@{}}
when was the last time \textbf{cleveland browns} were in the playoffs\\ 
has any \textbf{nfl} team gone 16-0 and won the \textbf{superbowl}\\ 
what \textbf{nfl} team never went to the \textbf{super bowl}
\end{tabular} \\ 
\midrule
2 (Movies) & \begin{tabular}[c]{@{}l@{}}
who is the \textbf{actor} that \textbf{plays} spencer reid\\ 
who died in vampire diaries \textbf{season} 1 \textbf{episode} 17\\
who \textbf{played} the mother on father knows best\\
how many \textbf{season} of the waltons are there
\end{tabular} \\ 
\midrule
3 (Human biology) & \begin{tabular}[c]{@{}l@{}}
where does the \textbf{amino acid} attach to \textbf{trna}\\ 
when a person is at rest approximately how much \textbf{blood} is being held within the \textbf{veins}\\ 
where are \textbf{lipids} synthesized outside of the endomembrane system\end{tabular} \\ 
\midrule
4 (Origin)    & \begin{tabular}[c]{@{}l@{}}
where does the last name broome \textbf{come from}\\ 
where does spring water found in mountains \textbf{come from}\\ 
where does the last name de leon \textbf{come from}
\end{tabular} \\ 
\midrule
5 (Soccer)    & \begin{tabular}[c]{@{}l@{} }
who is going to the 2018 \textbf{world cup}\\
when did \textbf{fifa} first begun and which country was it played \\ 
who came second in the \textbf{world cup} 2018
\end{tabular} \\ 
\midrule
6 (War) & \begin{tabular}[c]{@{}l@{}}
who fought on the western front during \textbf{ww1}\\ 
when did the allies start to win \textbf{ww2}\\ when did the united states declare \textbf{war} on \textbf{germany}\\ 
what were the most effective weapons in \textbf{ww1} 
\end{tabular}\\ 
\midrule
7 (Geography) & \begin{tabular}[c]{@{}l@{}}
how wide is the \textbf{mississippi} river at davenport \textbf{iowa}\\ 
bob dylan musical girl from the north \textbf{country} \\ 
what are the four \textbf{deserts} in \textbf{north america}
\end{tabular}\\ 
\bottomrule
\end{tabular}
}
\end{table*}

%% file: fig-prompting-analysis.tex
\begin{figure}[!ht]
        \centering  
        \includegraphics[width=0.45\linewidth]{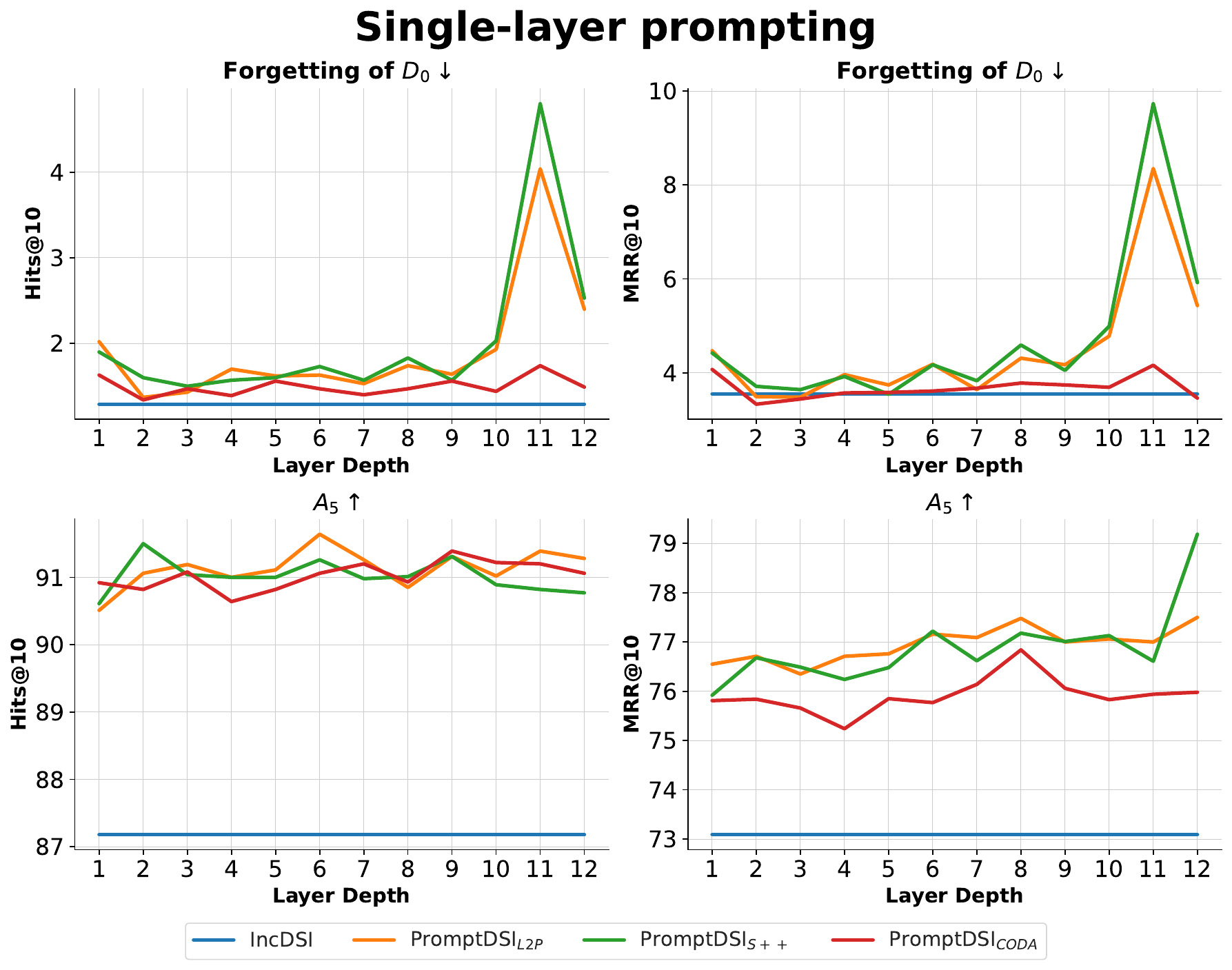}
        \rulesep
        \includegraphics[width=0.45\linewidth]{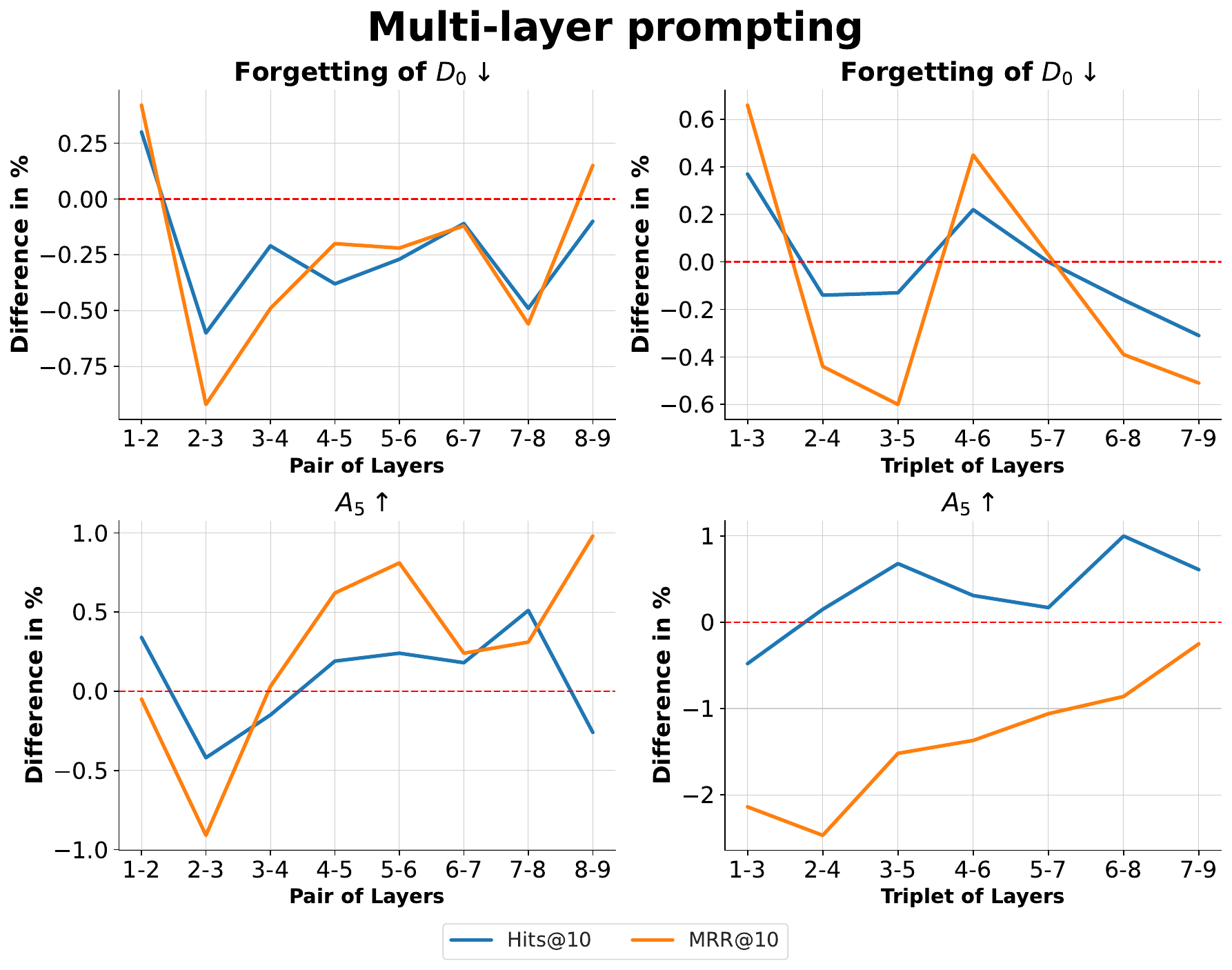}
        \caption{Layer-wise prompting analysis of SBERT-based PromptDSI on NQ320k.  
        \textbf{Left half:} Single-layer prompting analysis of different PromptDSI variants.  
        \textbf{Right half:} Multi-layer prompting analysis of SBERT-based PromptDSI$_\text{S++}$, illustrating the performance changes (i.e., percentage gains and drops) when using 2-layer prompting (first column) and 3-layer prompting (second column) compared to single-layer prompting at layer 1, indicated by the red dashed line.
        }
    \label{fig:sbert_single_layer_study}
\end{figure}

%% file: 6_conclusion.tex
\section{Conclusion}
We present PromptDSI, a novel prompt-based approach for rehearsal-free continual learning in document retrieval. We adapt prompt-based continual learning (PCL) methods and introduce single-pass PCL. We validate our approach on three standard PCL methods. Our topic-aware prompt pool addresses parameter underutilization, ensuring diverse and efficient prompt usage, while also enhancing interpretability. Results show that PromptDSI outperforms rehearsal-based DSI++ and matches IncDSI in mitigating forgetting, while achieving superior performance on new corpora.